\documentclass[doublecol]{epl2} 

\usepackage{amsmath,amssymb}

\usepackage{graphicx,color}

\usepackage{amsthm}

\usepackage[colorlinks=true]{hyperref}

\usepackage[T1]{fontenc}

\usepackage{bbm}

\newcommand{\ii}{\mathrm{i}}
\renewcommand{\vec}[1]{\mathbf{#1}}

\let \Im \relax
\DeclareMathOperator{\Im}{Im}
\let \Re \relax
\DeclareMathOperator{\Re}{Re}

\title{Cutting off the non-Hermitian boundary from an anomalous Floquet topological insulator}

\author{Bastian H{\"o}ckendorf\thanks{bastian.hoeckendorf@uni-greifswald.de} \and Andreas Alvermann \and Holger Fehske}
\shortauthor{Bastian H{\"o}ckendorf \etal}

\institute{                    
  Institut f{\"u}r Physik, Universit{\"a}t Greifswald, Felix-Hausdorff-Str. 6, 17489 Greifswald, Germany
}

\abstract{In two-dimensional anomalous Floquet insulators, non-Hermitian boundary state engineering can be used to completely separate chiral boundary states from bulk bands in the quasienergy spectrum.
The topological properties of such spectrally separated boundary states are no longer restricted by the strict bulk-boundary correspondence of Hermitian systems. 
We show that this additional topological freedom enables one 
to faithfully transfer the topological properties of a boundary attached to a Floquet insulator
to a non-Hermitian Floquet chain obtained by physically cutting off the boundary from the bulk.
We implement this scenario for a simple model of an anomalous Floquet insulator with Hermitian and non-Hermitian boundaries, and discuss the relevance of our construction for the experimental realization of non-Hermitian topological phases that connect dimensions one and two.
}

\begin{document}

\maketitle

The cornerstone of topological band theory is the bulk-boundary correspondence which provides a fundamental connection between the insulating bulk and chiral boundaries of a topological insulator~\cite{HasanKane2010, RevModPhys.83.1057, RevModPhys.88.035005, RevModPhys.91.015006, KaneMelePRL, Konig2007}. 
According to the correspondence, chiral boundary states exist 
if and only if the boundary is attached to a topologically non-trivial bulk. Since their existence is exclusively determined by the bulk topology, the boundary states are impervious to boundary deformations. This provides robustness but also serves as a fundamental restriction, in the following way: 
Suppose we try to extract a chiral boundary state from a topological insulator by cutting off one of its boundaries. Any such attempt necessarily fails, because the chiral boundary state moves towards the newly created boundary of the topological insulator. The cut off boundary, which is no longer attached to a topologically non-trivial bulk, does not possess chiral states.

In this work, we demonstrate that a \emph{non-Hermitian} boundary of a two-dimensional anomalous Floquet insulator can retain its topological properties when it is cut off from the bulk, provided that the boundary states are spectrally separated from the bulk bands.
To explain this scenario let us briefly review its three essential ingredients: the topological phases of anomalous (Hermitian) Floquet insulators~\cite{Rudner}, the topological phases of non-Hermitian Floquet chains~\cite{hckendorf2019nonhermitian, fedorova2019topological}, and the use of non-Hermitian boundary state engineering (BSE)~\cite{PhysRevLett.123.190403}.

In anomalous Floquet insulators~\cite{KitagawaPRB, Rudner, HockendorfJPA, Nathan, HockendorfPRB, PhysRevB.96.155118, PhysRevX.6.021013, HAF19, Maczewsky, Mukherjee, Peng2016, GreifRostock},
 non-trivial topology emerges through the dynamical evolution of the system even though each of the individual Floquet bands is topologically trivial. This situation is associated with a non-zero value of the bulk invariant~\cite{Rudner}
\begin{equation}
\label{W3inv}
\begin{aligned}
W_3 &=\frac{1}{8\pi^2} \int_0^T  \iint_{\mathcal{B}} \\ &\mathrm{tr}\Big(U^{-1}\partial_{t}U \big[U^{-1}\partial_{k_x}U, U^{-1}\partial_{k_y}U  \big] \Big) \mathrm{d}k_x \mathrm{d}k_y \mathrm{d}t \; .
\end{aligned}
\end{equation}
Here, $U= U(\vec k, t)$ is the Floquet-Bloch propagator, given as the solution of the Schr{\"o}dinger equation $\ii\partial_t U(\vec k, t)=H(\vec k, t) U(\vec k,t)$ for a time-periodic Hamiltonian $H(\vec k, t)=H(\vec k, t+T)$, and $\mathcal{B}$ denotes the Brillouin zone of momenta $\vec k$. Formally, the $W_3$ invariant is a $\mathbb Z$-valued topological invariant only if the Floquet-Bloch propagator fulfills the constraint $U(\vec k,T)=\mathbbm 1$, that is for flat Floquet bulk bands.
However, any propagator with a gap can be deformed continuously until this constraint is fulfilled, without changing the topology~\cite{Rudner}.

\begin{figure}
\includegraphics[width=1\columnwidth]{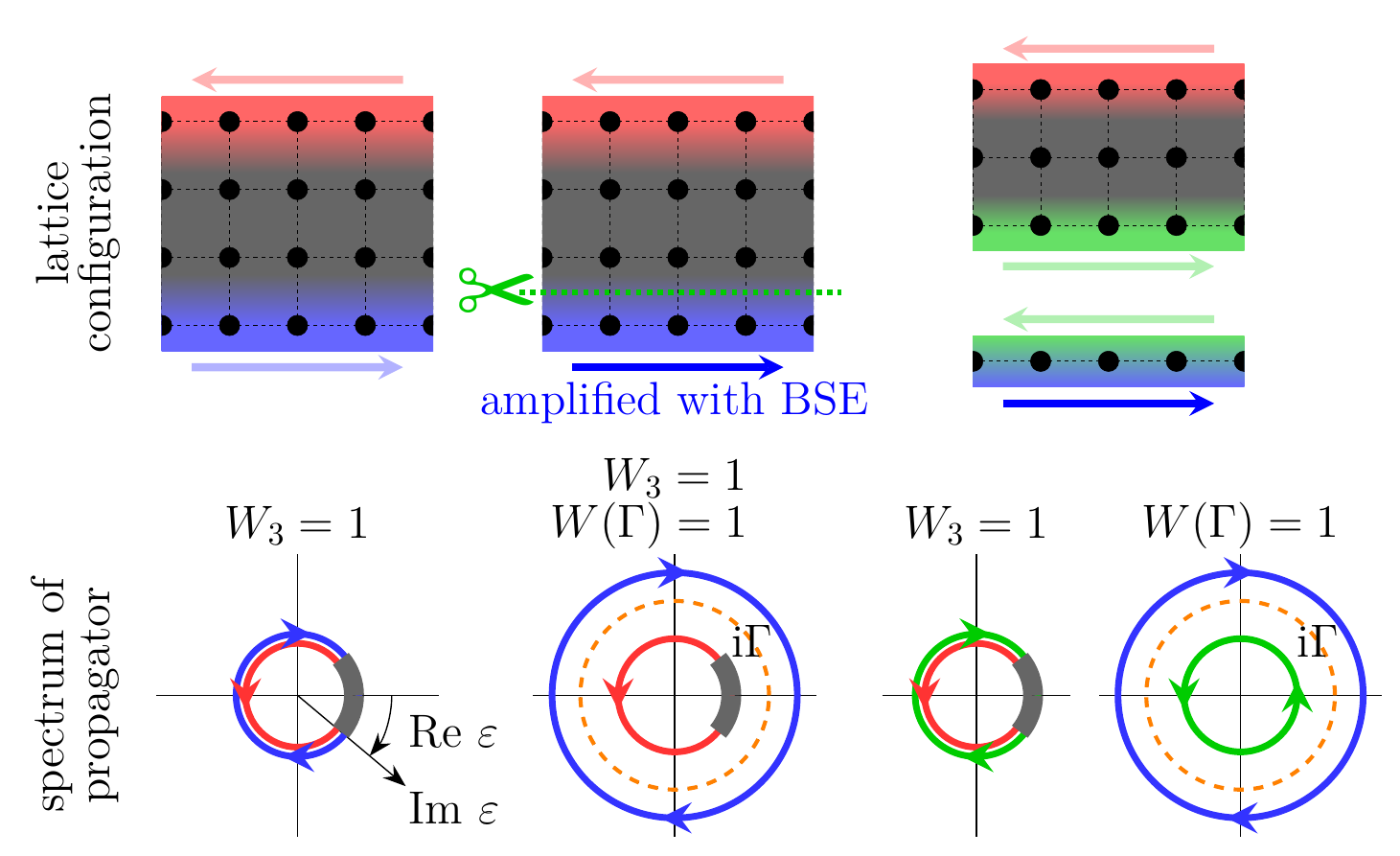}
\caption{Top row: Blueprint for the construction of a one-dimensional non-Hermitian Floquet chain with non-zero winding number $W(\Gamma)$ from the non-Hermitian boundary of an anomalous Floquet insulator with non-zero $W_3$ invariant. 
Bottom row: Eigenvalues $e^{-\ii\varepsilon(k)}$ of the Floquet-Bloch propagator $\hat U(k)$ for the different configurations. Real (imaginary) parts of the quasienergy $\varepsilon$ correspond to the phase (magnitude) of the eigenvalues $e^{-\ii\varepsilon(k)}$, as sketched in the left panel. Colors indicate the spatial position of the corresponding eigenstates, e.g. bulk bands are indicated by thick gray arcs corresponding to the gray bulk in the top row.
Arrows specify the direction of increasing momentum in the dispersion $k \mapsto e^{-\ii \varepsilon(k)}$.
The third (fourth) panel shows the spectrum of the remaining bulk (the resulting chain) after cutting off the bottom boundary.
}
\label{Fig:1}
\end{figure}

The immediate signature of the anomalous topological phase is the appearance of chiral boundary states in the spectrum of the Floquet-Bloch propagator $\hat U(k)\equiv U(k,T)$ on a semi-infinite strip, using the geometry in the top row of Fig.~\ref{Fig:1}.
Here, $k \equiv k_x$ denotes the momentum parallel to the boundaries of the strip (oriented along the $x$-axis). The spectrum  $\{ e^{-\ii\varepsilon(k)}\}$ of $\hat U(k)$ with the real-valued Floquet quasienergies $\varepsilon(k)$ lies on the unit circle.
The boundary states form chiral loops $k \mapsto e^{-\ii \varepsilon(k)}$ that wind around the unit circle and connect opposite ends of the bulk bands (see the first panel in the bottom row of Fig.~\ref{Fig:1}).
Since the loops are attached to the bulk bands they are non-contractible and topologically non-trivial.

 Especially for flat bulk bands, that is $U(\vec k,T)=\mathbbm 1$ as in the definition of the $W_3$ invariant,
 the loops wind fully around the unit circle.
 We can thus assign a winding number to each boundary, say
 \begin{subequations}
 \label{WindingHermitian}
 \begin{align}
 W
&=\frac{\ii}{2\pi} \sum_{j \in \mathrm{bottom}} \,  \int_{-\pi}^{\pi}   e^{\ii \varepsilon_j(k)} \,\partial_k \mkern1mu e^{-\ii \varepsilon_j(k)} \, \mathrm{d}k  \\
&=\frac{1}{2\pi} \sum_{j \in \mathrm{bottom}}  \varepsilon_j(k+2\pi) - \varepsilon_j(k)
\end{align}
\end{subequations}
for the bottom boundary of the strip in Fig.~\ref{Fig:1}.
Since the top and bottom boundary are spatially separated by the bulk, only states on one boundary are counted in the winding number.
The bulk-boundary correspondence states that $W = W_3$, independently of the boundary.
On the bottom boundary, $W = W_3 > 0$ corresponds to boundary states that move to the right (as in Fig.~\ref{Fig:1}).

Non-Hermitian Floquet chains are another system in which non-trivial topological phases are associated with non-contractible loops~\cite{hckendorf2019nonhermitian, fedorova2019topological}. 
The topological states of these chains appear as non-contractible loops with opposite chirality in the now complex-valued quasienergy spectrum, in analogy to the boundary states of anomalous Floquet insulators.
Here, however, chiral states are not spatially separated by a bulk.
Instead, the non-contractible loops associated with chiral states are spectrally separated by an imaginary gap $\ii \Gamma$ (see the last panel in the bottom row of Fig.~\ref{Fig:1}).
For the complex quasienergy $\varepsilon(k)=\Re \varepsilon(k)+\ii \Im \varepsilon(k)$, we can distinguish between states above ($\Im \varepsilon(k) > \Gamma$) and below ($\Im \varepsilon(k) < \Gamma$) the imaginary gap.
The topological phase is determined by the number of non-contractible loops above the imaginary gap,
and thus classified by the $\mathbb Z$-valued winding number~\cite{hckendorf2019nonhermitian}
\begin{equation} \label{W1}
W(\Gamma)
=\frac{\ii}{2\pi} \sum_{e^\Gamma < |e^{-\ii \varepsilon_j}|} \,  \int_{-\pi}^{\pi}   e^{\ii \varepsilon_j(k)} \,\partial_k \mkern1mu e^{-\ii \varepsilon_j(k)} \, \mathrm{d}k \; ,
\end{equation}
which counts the net chirality of these loops. 
By definition, the winding number can only change when the imaginary gap closes.

The winding number $W(\Gamma)$ can be viewed as a generalization of the winding number $W$ for a Hermitian system in Eq.~\eqref{WindingHermitian}.
It is tempting to relate the two invariants in the form of an equation ``$W = W(-\infty)$'', recognizing that the Hermitian system has the trivial imaginary gap $\Gamma=-\infty$.
There is, however, an important distinction to be made.
In a two-dimensional Floquet insulator chiral boundary states appear on opposite boundaries with the same chirality, thus moving either left or right, and are spatially separated by the bulk. The winding number $W$ counts the net chirality of states on one boundary.
If the left and right moving chiral states were to appear on the same boundary, they would cancel each other (unless additional symmetries are present~\cite{HAF19,Maczewsky}).
In a one-dimensional non-Hermitian Floquet chain, where there is no bulk, chiral states are not spatially separated.
To avoid cancellation, these states have to be separated spectrally by an imaginary gap $\ii \Gamma$.
Indeed, for the trivial imaginary gap $\Gamma=-\infty$ we obtain the total chirality $W(-\infty)=0$  of the spectrum, which is necessarily zero due to the invertibility of the propagator~\cite{hckendorf2019nonhermitian}.

Non-Hermitian BSE~\cite{PhysRevLett.123.190403} describes a process in which non-Hermiticity is introduced locally at the boundaries of an anomalous Floquet insulator, with the goal to spectrally separate chiral boundary states from the bulk bands. In this way, BSE allows for deliberate manipulation of chiral boundary states and topological boundary transport.
Here, we will show that non-Hermitian BSE establishes a connection between anomalous Floquet insulators with $W_3 \ne 0$, and non-Hermitian Floquet chains with $W(\Gamma) \ne 0$. 

To understand this connection, consider, exemplarily, the configuration in the central panel of the top row of Fig.~\ref{Fig:1} where BSE is applied to the bottom boundary of an anomalous Floquet insulator. 
Since after BSE an imaginary gap separates the boundary state on the bottom boundary from the rest of the quasienergy spectrum, the two-dimensional semi-infinite strip is no longer described just by a non-zero $W_3$ invariant, but also by a non-zero winding number $W(\Gamma)$.

To ``cut off'' the boundary, we set all couplings between the non-Hermitian boundary and the bulk to zero, as in the right panel of the top row of Fig.~\ref{Fig:1}.
This creates a new Hermitian boundary at the bottom of the anomalous Floquet insulator, and a separated non-Hermitian Floquet chain.
The topological result of this procedure is determined by the behavior of the two invariants $W_3, W(\Gamma)$.
First, the bulk invariant $W_3$ is not affected by the removal or addition of a boundary.
We conclude that
a chiral boundary state appears on the new Hermitian boundary of the anomalous Floquet insulator, in accordance with the bulk-boundary correspondence $W=W_3$. At the same time, a second state with opposite propagation direction is created on the new non-Hermitian Floquet chain since the total winding number $W(-\infty)$ must remain zero. As long as the imaginary gap stays open while cutting, the second state is
separated from the original boundary state by an imaginary gap,
and the winding number $W(\Gamma)$ can not change and remains non-zero for the new chain.

In the interpretation given here, we read Fig.~\ref{Fig:1} from left to right:
We start with an anomalous Floquet insulator ($W=W_3 \ne 0$), spectrally separate chiral boundary states with non-Hermitian BSE (which achieves $W(\Gamma) = W \ne 0$), cut off the non-Hermitian boundary, and obtain a  Floquet chain with non-trivial topology since $W(\Gamma) \ne 0$ does not change during the cut. The remaining bulk system is still an anomalous Floquet insulator, since $W_3$ does not change during the cut.

Alternatively, we can read Fig.~\ref{Fig:1} from right to left. Now, we start with a non-Hermitian Floquet chain that possesses non-trivial topology  if chiral loops are separated by an imaginary gap ($W(\Gamma) \ne 0$), and attach the chain to an anomalous Floquet insulator with chiral states on opposite boundaries ($W=W_3 \ne 0$).
For infinitesimal coupling between the chain and the bulk, states on the new non-Hermitian boundary of the joint chain-bulk system are described separately by the $W_3$ invariant and the $W(\Gamma)$ winding number.
When we gradually increase the coupling and reduce non-Hermiticity, changing them back to the values of the homogeneous system,  two things can happen.
If we start with $W_3 = W(\Gamma)$, the chiral states of the chain and on the boundary of the Floquet insulator are compatible. Then, we essentially perform inverse BSE. The states above the imaginary gap survive, and the system evolves continuously towards the anomalous Floquet insulator in the leftmost panel in Fig.~\ref{Fig:1}.
But if we start with $W_3 \ne W(\Gamma)$, the chiral states of the chain above the imaginary gap are incompatible with the chiral boundary states of the anomalous Floquet insulator. Therefore, the imaginary gap has to close during the procedure, allowing $W(\Gamma)$ to change its value
such that we finally recover the anomalous Floquet insulator with the correct net chirality $W = W_3$ of boundary states on the bottom boundary.

\begin{figure}
\includegraphics[width=1\columnwidth]{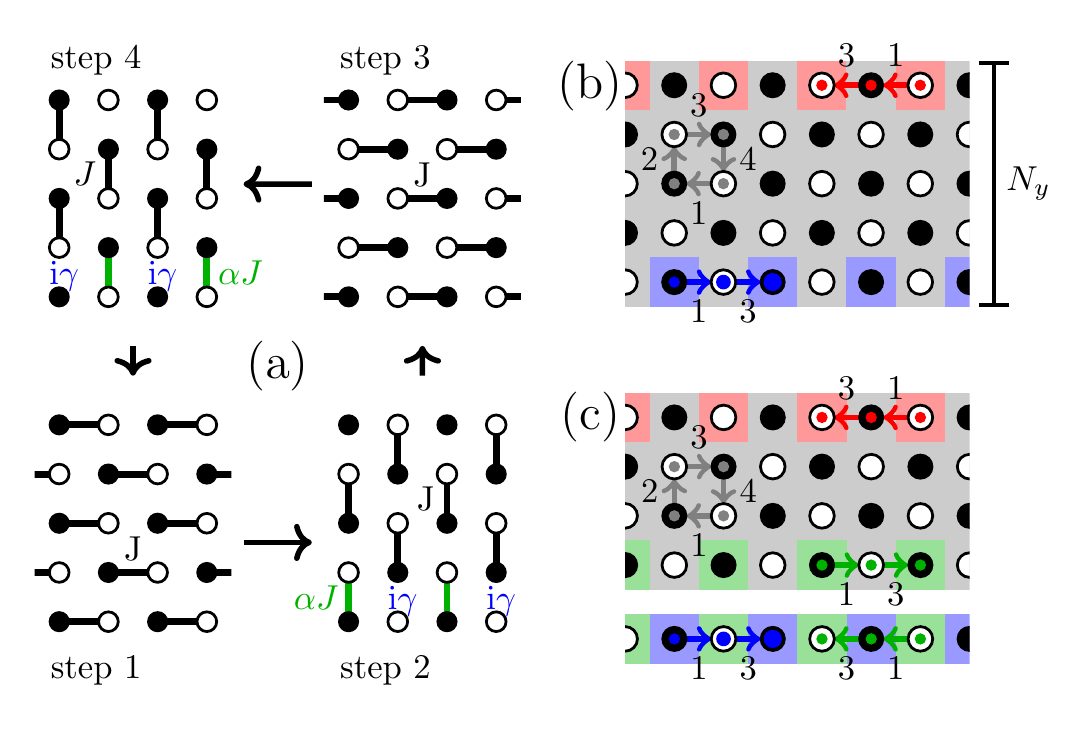}
\caption{
(a) Driving protocol for an anomalous Floquet insulator on a semi-infinite strip along the $x$-axis, with pairwise couplings between adjacent lattice sites (filled and open circles).  Panel (b) [panel (c)] shows the patterns of motion during one cycle of the driving protocol for perfect coupling, $\gamma>0$ and $\alpha=1$ ($\alpha=0$). 
}
\label{Fig:2}
\end{figure}

To present an explicit example, we now apply this procedure to the standard model~\cite{Rudner} of an anomalous Floquet insulator which is sketched in Fig.~\ref{Fig:2}(a). The $2+1$-dimensional driving protocol is implemented on a square lattice with lattice sites $\textcolor{black}{\bullet}$ (filled circles) and $\textcolor{black}{\circ}$ (open circles). We enforce translational invariance along the $x$-axis and open boundary conditions along the $y$-axis such that the lattice possesses a top and bottom boundary. The number of sites between the two boundaries is indexed by the parameter $N_y$  [see Fig.~\ref{Fig:2}(b)].
To describe a two-dimensional Floquet insulator we use large $N_y \gg 1$, for a one-dimensional chain we will set $N_y = 1$.
 The time-periodic Hamiltonian $H(k,t)$
 of the driving protocol, with momentum $k$ along the $x$-axis, cycles through four consecutive steps $H^{(1)}(k), ..., H^{(4)}(k)$ of duration $\delta t=T/4$, 
each of which couples two different adjacent sites of the lattice. The propagators of each step are $U^{(i)}(k)=\exp(-\ii \delta t H^{(i)}(k))$ and the Floquet propagator of the full driving protocol is given by $\hat U(k)=U^{(4)}(k)\cdot \cdot  \cdot U^{(1)}(k)$.

To specify the Hamiltonians of the individual steps, we use standard bra-ket notation, where $|\textcolor{black}{\bullet},n_y \rangle$  ($|\textcolor{black}{\circ},n_y \rangle$) denotes the Bloch state of a particle on the $\textcolor{black}{\bullet}$ ($\textcolor{black}{\circ}$) sites in the $n_y$-th layer between the bottom and top boundary with $n_y=1, ..., N_y$. To avoid notational clutter, we omit the explicit momentum dependence $|\textcolor{black}{\bullet},n_y \rangle \equiv |\textcolor{black}{\bullet},n_y,k \rangle$,  $|\textcolor{black}{\circ},n_y \rangle \equiv |\textcolor{black}{\circ},n_y,k \rangle$ of the Bloch states. With these conventions, the four steps are expressed as
\begin{subequations}
\begin{multline}\label{eq:step1}
 H^{(1)}(k) = 
 \\
 \frac{J}{\delta t} \sum_{n_y=1}^{N_y} \Big(e^{-\ii k/2}|\textcolor{black}{\circ},n_y \rangle \langle\textcolor{black}{\bullet},n_y |+ e^{\ii k/2}|\textcolor{black}{\bullet},n_y \rangle \langle \textcolor{black}{\circ},n_y |  \Big) \; ,
\end{multline}
\begin{multline}\label{eq:step2}
 H^{(2)}(k) = \frac{J}{\delta t} \sum_{n_y=3}^{N_y} \Big(|\textcolor{black}{\circ},n_y \rangle \langle\textcolor{black}{\bullet},n_y-1 |+ |\textcolor{black}{\bullet},n_y-1 \rangle \langle \textcolor{black}{\circ},n_y |  \Big) 
\\
+\frac{\alpha J}{\delta t}  \Big(|\textcolor{black}{\circ},2 \rangle \langle\textcolor{black}{\bullet},1 |+ |\textcolor{black}{\bullet},1 \rangle \langle \textcolor{black}{\circ},2 |  \Big) +\ii \frac{\gamma}{\delta t} |\textcolor{black}{\circ},1 \rangle \langle\textcolor{black}{\circ},1 |
 \; ,
 \end{multline}
\begin{equation}\label{eq:step3}
 H^{(3)}(k) = S(k) H^{(1)}(k) S^{-1}(k) \; ,
\end{equation}
\begin{equation}\label{eq:step4}
 H^{(4)}(k) =S(k) H^{(2)}(k) S^{-1}(k) \; .
\end{equation}
\end{subequations}
The symmetry operator $S(k)=\sum_{n_y=1}^{N_y} (|\textcolor{black}{\circ},n_y \rangle \langle\textcolor{black}{\bullet},n_y|+|\textcolor{black}{\bullet},n_y \rangle \langle\textcolor{black}{\circ},n_y|)$ exchanges the two types of sites. Note that we normalize $k$, with the Brillouin zone given by the interval $k\in [-\pi,\pi)$.
The three parameters $J$, $\gamma$, and $\alpha$ control the coupling strength in the bulk, the strength of non-Hermitian BSE on the bottom boundary, and the coupling of the bottom boundary to the bulk.

We first focus on the case $\alpha=1$, where the bottom boundary is fully attached to the rest of the lattice.
The pairwise coupling between two sites in each of the four steps is effectively described by the $2\times2$ Hermitian Hamiltonian
\begin{equation}
\label{eq:H^u}
H_{J}=\frac{J}{\delta t} \begin{pmatrix}0 & 1 \\ 1 & 0 \end{pmatrix}=\frac{J}{\delta t} \sigma_x \; .
\end{equation}    
The associated propagator $U_J=\exp(-\ii \delta t H_J)=\cos(J)-\ii  \sin(J) \sigma_x$, of a time step, is a periodic function of $J$. Since $U_{J}=(-1)^m U_{J+m\pi}$ for every $m\in \mathbb Z$, we may restrict ourselves to the parameter range $J\in[0,\pi)$. For $0\le J<\pi/4$ and $3\pi/4< J< \pi$, the $W_3$ invariant vanishes, which indicates that the system is topologically trivial.  On the other hand, the $W_3$ invariant is non-zero for $\pi/4<J<3\pi/4$. This indicates that the system is in the anomalous Floquet topological phase, which makes it a candidate for BSE.

The simplest way to facilitate BSE is to implement gain or loss on boundary sites that
do not couple to other sites in the respective step of the protocol~\cite{PhysRevLett.123.190403}. Here, we place a non-Hermitian imaginary potential $\ii \gamma/\delta t$ onto the isolated sites in steps $2$ [see Eq.~\eqref{eq:step2}] and $4$ [see Eq.~\eqref{eq:step4}] at the bottom boundary. For $\gamma>0$, the imaginary potential corresponds to gain and for $\gamma<0$ to loss.
Note that the rest of the lattice, including the top boundary, is Hermitian also for $\gamma\ne 0$. Therefore, Eq.~\eqref{W3inv} remains the appropriate bulk invariant, and generalizations of the $W_3$ invariant to non-Hermitian systems~\cite{li2020twodimensional} are not required.

For the special value $J=\pi/2$, which we call perfect coupling, we have $U_J=-\ii\sigma_x$. A full amplitude transfer occurs between coupled sites and the driving protocol enforces the trajectories shown in Fig.~\ref{Fig:2}(b). An excitation in the bulk moves in a closed loop, while an excitation starting on a $\textcolor{black}{\bullet}$ site ($\textcolor{black}{\circ}$ site) at the bottom (top) boundary is transported by two sites to the right (left).  This pattern of motion gives rise to chiral boundary states on the bottom and top boundary. The chiral boundary state on the bottom boundary gets amplified (attenuated) by a factor $e^{2\gamma}$ for $\gamma>0$ ($\gamma<0$). All other excitations have constant amplitude. This corresponds to the quasienergy dispersions $\varepsilon(k)=0$ for the bulk band and $\varepsilon(k)=\pi+k+2\ii \gamma$ [$\varepsilon(k)=\pi-k$] for the chiral boundary state on the bottom [top] boundary. Therefore, we have $W(\Gamma)=1$ for $\gamma>0$ and $W(\Gamma)=-1$ for $\gamma<0$ with the imaginary gap $\ii \Gamma$ at $\Gamma=\gamma$.

\begin{figure}
\hspace*{\fill}
\includegraphics[width=0.99\columnwidth]{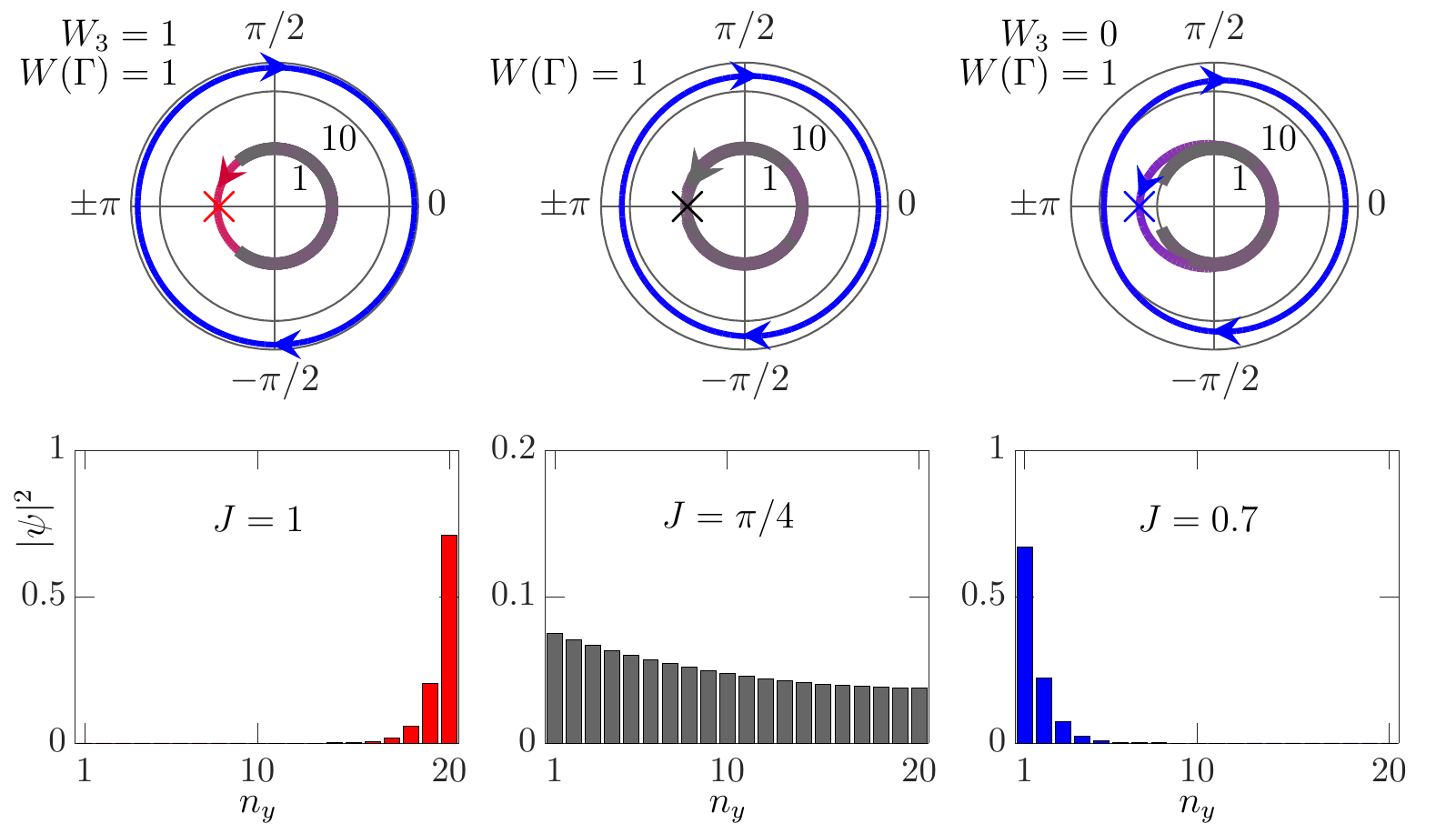}
\hspace*{\fill}
\caption{Top row: Spectrum of the Floquet-Bloch propagator for the driving protocol in Fig.~\ref{Fig:2} with $N_y=20$ sites between the bottom and top boundary and $\alpha=1$, $\gamma=1.8$. The bulk phase transition (central panel, $J=\pi/4$) separates the phase with a non-trivial (left panel, $J=1 > \pi/4$) and trivial bulk (right panel, $J=0.7 < \pi/4$).
Note that we use a logarithmic radial axis in all panels. 
The color of the curves indicates the amplitude distribution of the eigenfunctions on the red, blue, and gray colored area of the lattice in Fig.~\ref{Fig:2}(b).
Bottom row: Amplitude distribution of the eigenfunctions $\psi(k)$ of $\hat U(k)$ at $k=0$ as a function of $n_y$ for the eigenvalues marked with an $\times$ in the top row.
}
\label{Fig:3}
\end{figure}

For sufficiently large values of $|\gamma|$, the imaginary gap stays open for any finite non-perfect coupling ($J \not\in \{0, \pi/2\}$).
Therefore, the boundary state on the bottom boundary remains spectrally separated from the rest of the spectrum
and exists independently of the value of the $W_3$ bulk invariant.
Figure~\ref{Fig:3} demonstrates this phenomenon. At $J=1$ (left panels), the $W_3$ invariant and winding number $W(\Gamma)$ are both non-zero. 
Increasing $J$,  the topological phase transition occurs at $J=\pi/4$ (middle panels).
Passing through the transition, the $W_3$ invariant changes to zero,
such that the boundary state on the Hermitian top boundary, which strictly adheres to the bulk-correspondence, disappears.
The imaginary gap, however, stays open, which means that the winding number $W(\Gamma)$ does not change. Consequently, the boundary state on the non-Hermitian bottom boundary can not disappear and persists even beyond the topological phase transition at $J=0.7$ (right panels).
This implies the (partial) breakdown of the bulk-boundary correspondence.

The breakdown of the bulk-boundary correspondence has previously been observed in many different non-Hermitian static systems~\cite{PhysRevX.8.031079, PhysRevB.99.235112, PhysRevX.9.041015, PhysRevLett.121.086803, PhysRevLett.121.136802, PhysRevB.99.201103, PhysRevLett.121.026808, 2019arXiv190711562H}.  
In these system, the breakdown is a consequence of the non-Hermitian skin effect, in which bulk states become exponentially localized when open boundary conditions are introduced.
The non-Hermitian skin effect is thus associated with a significant change in the bulk energy spectrum. In our example of an anomalous Floquet insulator, the bulk quasienergies do not depend on the boundary condition. Here, the breakdown of the bulk-boundary correspondence is not a consequence of the non-Hermitian skin effect, but enabled by the spectral separation between bulk and boundary states induced through BSE.  Indeed, we observe that the conventional bulk-boundary correspondence is immediately recovered for the remaining Hermitian bulk when the non-Hermitian boundary is cut off.

To summarize, we have the following situation:
The boundary state on the Hermitian top boundary exists only for a topologically non-trivial bulk ($W = W_3 \ne 0$), in accordance with the bulk-boundary correspondence.
The boundary state on the non-Hermitian bottom boundary above the imaginary gap can coexist with a topologically trivial bulk ($W(\Gamma) \ne 0, W_3 = 0$), and thus violates a potential non-Hermitian bulk-boundary correspondence.
Note that for $W_3 = 0$, chiral states with $W(\Gamma) \ne 0$ have to appear in pairs of opposite chirality on the bottom boundary (see the bottom row of Fig.~\ref{Fig:3}), since the total winding number $W(-\infty)$ is always zero but the bulk-boundary correspondence prohibits that the chiral states are placed separately on the top and bottom boundary. %
Here, we  directly observe that the bottom boundary is essentially a non-Hermitian Floquet chain, with non-zero winding number $W(\Gamma) \ne 0$, that is attached to a trivial insulator with $W_3=0$.
Once again we stress that two states of opposite chirality would cancel each other on a Hermitian boundary, in agreement with the trivial bulk topology, but in our case the imaginary gap prevents cancellation.

Now we want to fully cut off the non-Hermitian boundary, and thus set $\alpha=0$ in the model.
After the cut, the former bottom boundary can be regarded as a separate one-dimensional chain. 
For perfect coupling, the trajectories of states on the chain have already been shown in Fig.~\ref{Fig:2}(c). The two counterpropagating states of the chain have the same trajectories and quasienergies as the boundary states in Fig.~\ref{Fig:2}(b). Again, we have $W(\Gamma)=1$ for $\gamma>0$ and $W(\Gamma)=-1$ for $\gamma<0$ with the imaginary gap $\ii \Gamma$ at $\Gamma=\gamma$. Note that there is no spatial overlap between the wave functions of the two bulk states for perfect coupling. 

\begin{figure}
\hspace*{\fill}
\includegraphics[width=0.99\columnwidth]{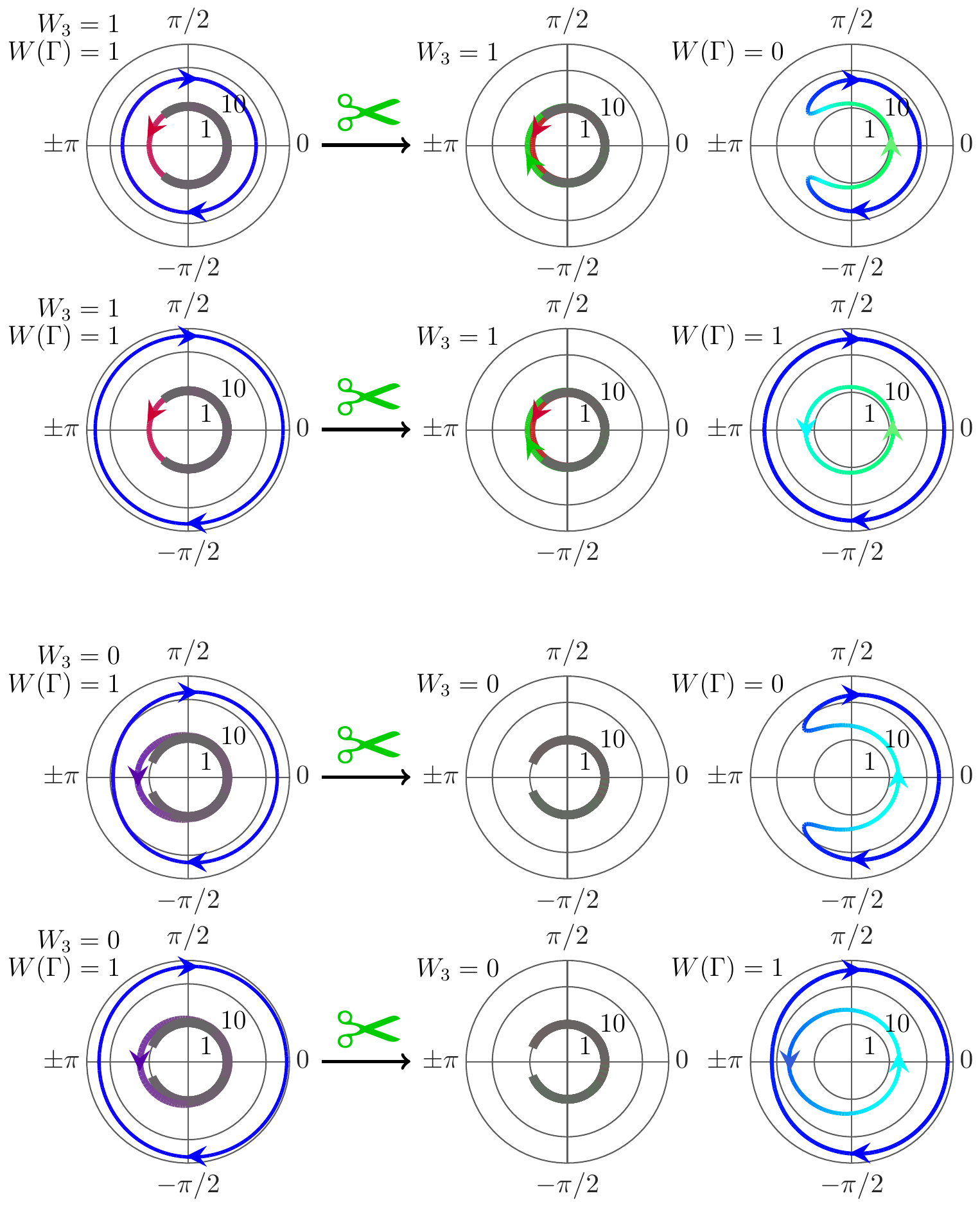}
\hspace*{\fill}
\caption{Same as Fig.~\ref{Fig:3}, now showing the spectra before ($\alpha=1$, left panels) and after ($\alpha=0$, central and right panels) the bottom boundary has been cut off. In the top row we use the parameters $J=\gamma=1$, in the second row we use $J=1$, $\gamma=1.8$, in the third row we use $J=0.7$, $\gamma=1.8$, and in the bottom row we use $J=0.7$, $\gamma=2.1$. In all panels, we have $N_y=20$. The color of the curves now indicates the amplitude distribution of the eigenfunctions on the red, blue, gray, and green colored area of the lattice in Fig.~\ref{Fig:2}(c).
}
\label{Fig:4}
\end{figure}

For non-perfect coupling, this is no longer the case. Therefore, the imaginary gap only persists for sufficiently large values of $|\gamma|$.
The critical value $\gamma_c$, at which the imaginary gap closes, follows from the eigenvalues of $\hat U(k)$ for the chain, analogous to Ref.~\cite{hckendorf2019nonhermitian}. The two eigenvalues are
\begin{equation}
\label{eq:Fl_eigenvalues}
\begin{aligned}
e^{-\ii \varepsilon_{1,2}(k)}  =   
 e^{\gamma} \Big(& 1-2 \sin^2 J\cos^2 \kappa
\\ 
& \pm 2 \sin J  \cos \kappa \sqrt{\sin^2 J \cos^2 \kappa-1}\Big)  \; ,
\end{aligned}
\end{equation}
with $\kappa=k/2+\ii \gamma/2$. 
The topological phase transition between a trivial chain with $W(\Gamma)=0$ and a non-trivial chain with $W(\Gamma)=1$ occurs when the square root in Eq.~\eqref{eq:Fl_eigenvalues} vanishes, which happens at $\gamma = \pm \gamma_c$ for
$\gamma_c(J)= 2~\mathrm{arcosh}(1/\sin |J|)$.

Note that the critical value $\gamma_c$ is generally larger than the value of $\gamma$ at which the boundary state becomes spectrally separated from the rest of the spectrum for $\alpha=1$. This is due to the spatial overlap between the counterpropagating states in the chain for $\alpha=0$ which reduces the imaginary gap induced by BSE. The various scenarios that emerge in the present driving protocol after the bottom boundary has been cut off are shown in Fig.~\ref{Fig:4}. In all cases, the bottom boundary of the anomalous Floquet insulator satisfies the bulk-boundary correspondence $W=W_3$. Non-Hermitian Floquet chains with non-zero winding number $W(\Gamma)$ are generated if $\gamma$ is above the critical value $\gamma_c$ as in the second and fourth row of Fig.~\ref{Fig:4}.

In conclusion, non-Hermitian Floquet chains can be constructed by cutting off the non-Hermitian boundaries of two-dimensional anomalous Floquet insulators. Conversely, a non-Hermitian Floquet chain can be attached to an anomalous Floquet insulator to obtain a system that combines aspects of two-dimensional Hermitian and one-dimensional non-Hermitian topological phases.
The crucial aspect in this construction is the imaginary gap that separates boundary states from the bulk spectrum. The imaginary gap provides topological protection for the boundary states, irrespectively of the bulk topology. In this way, chiral boundary states can even coexist with a topologically trivial bulk, in direct violation of the conventional bulk-boundary correspondence for two-dimensional Hermitian systems.

 Our construction enables a straightforward experimental realization of non-Hermitian Floquet chains by exploiting existing experimental designs~\cite{Maczewsky, Mukherjee, Peng2016, GreifRostock} for anomalous Floquet insulators. The present model can be implemented in photonic waveguide lattices~\cite{Maczewsky, Mukherjee}. In that context, non-Hermitian BSE could be realized through waveguide bending~\cite{Weimann2016}.  

An interesting route for future research is the application of our procedure to symmetry-protected anomalous Floquet topological phases or systems with different spatial dimensionality~\cite{PhysRevB.96.155118}. We expect that by cutting off the boundaries from other anomalous Floquet insulators, e.g. those with fermionic time-reversal symmetry~\cite{GreifRostock, PhysRevLett.123.190403}, novel non-Hermitian Floquet topological phases will emerge, which may also possess interesting transport properties.

\bibliographystyle{eplbib}

\end{document}